\documentstyle[12pt,a4]{article}
\newcommand{\beq}{\begin{equation}}
\newcommand{\eeq}{\end{equation}}
\newcommand{\beqn}{\begin{eqnarray}}
\newcommand{\eeqn}{\end{eqnarray}}
\newcommand{\np}{Nucl.Phys. \underline}
\newcommand{\pl}{Phys.Letters \underline}
\newcommand{\pr}{Phys.Rev. \underline}
\newcommand{\prep}{Phys.Reports. \underline}
\newcommand{\mmsbar}{$m_{\overline{MS}}$}
\newcommand{\mpole}{m_{{\rm pole}}}
\newcommand{\lb}{\bar\Lambda}

\newcommand{\vslash}{\slash \hspace{-6.3pt} v}
\begin{document}
\pagestyle{empty}
\setcounter{page}{0}
\begin{flushright}
CERN-TH.7517/94 \\
SHEP 94/95-13 \\
\end{flushright}
\vskip 0.6in
\centerline{\bf{RENORMALONS AND THE HEAVY QUARK EFFECTIVE THEORY}}
\vskip 0.8cm
\centerline{\bf{ G. Martinelli$^{a,*}$ and
C.T. Sachrajda$^b$}}
\mbox{} \\
\centerline{$^a$ Theory Division, CERN, 1211 Geneva 23, Switzerland.}
\centerline{$^b$ Dep. of Physics, University of Southampton,}
\centerline{Southampton SO17 1BJ, U.K.}

\abstract{We propose a non-perturbative method for defining the
higher dimensional operators which appear in the Heavy Quark Effective
Theory (HQET), such that their matrix elements are free of renormalon
singularities, and diverge at most logarithmically with the
ultra-violet cut-off. Matrix elements of these operators can be
computed numerically in lattice simulations of the HQET. We illustrate
our procedures by presenting physical definitions of the  binding
energy ($\lb$) and of the kinetic energy (-$\lambda_1/2m_Q$) of the
heavy quark in a hadron. This allows us to define a ``subtracted pole
mass", whose inverse can be used as the expansion parameter in
applications of the HQET.}

\vskip 0.8cm
\begin{flushleft} CERN-TH.7517/94 \\ February 1995
\end{flushleft} \vskip 0.8cm

\centerline{$^*$ On leave of absence  from Dip. di Fisica, Universit\`a
degli Studi ``La Sapienza", Rome, Italy. } \vfill\eject

\pagestyle{empty}\clearpage \setcounter{page}{1}
\setcounter{footnote}{0}
\pagestyle{plain}

\section{Introduction}\label{sec:intro}

The Heavy Quark Effective Theory (HQET) has developed during the last few
years into a very useful tool for the study of strong interaction
effects in heavy quark physics \cite{nw}-\cite{georgi} (for a
comprehensive review see ref.\cite{neubert}). In this approach physical
quantities are studied systematically as series in inverse powers of
the mass of the heavy quark~\footnote{In $B\rightarrow D$ decays the
expansion may be performed simultaneously in the inverse powers of the
masses of both heavy quarks.}. In particular, local composite operators
of QCD, whose matrix elements contain the long-distance gluonic effects
in physical processes, are expanded in terms of the operators of the
HQET
\beq
O^{QCD} = \sum_{n,\alpha}\frac{C_{n,\alpha}(m_Q/\mu)}{m_Q^n}O^{HQET}_
{n,\alpha}(\mu)\, ,
\label{eq:ope}\eeq
where $m_Q$ is the mass of the heavy quark and $\mu$ is the
renormalisation scale used in defining the renormalised operators
$O^{HQET}_{n,\alpha}(\mu)$. For the moment we assume that the operators
$O^{HQET}_{n,\alpha}(\mu)$ are renormalised in some scheme based on the
dimensional regularisation of ultra-violet divergences, such as the
$\overline{MS}$ scheme. The QCD operators $O^{QCD}$ generally also
depend on some renormalisation scale, which we take to be different
from $\mu$, and we do not exhibit this dependence explicitly.
Renormalon singularities lead to ambiguities in the Wilson coefficient
functions $C_{n,\alpha}$, and in the matrix  elements of the operators
$O_{n,\alpha}^{HQET}$. Although these ambiguities cancel (see
ref.\cite{muelleraachen} and references therein, and
refs.\cite{bb}-\cite{manohar}), the presence of renormalons requires an
alternative definition of the renormalised operators
$O_{n,\alpha}^{HQET}$, if the HQET is to be applicable beyond leading
order in the heavy quark mass, and in particular if the coefficient
functions are to be calculable using perturbation theory. The renormalon
singularities and the corresponding ambiguities which we are
considering here are those induced by the introduction of the expansion
in eq.(\ref{eq:ope}). Of course the matrix elements of the QCD operator
$O^{QCD}$ will themselves contain non-perturbative long-distance
effects, but, as usual,  these effects do not appear in the coefficient
functions, but only in the matrix elements of the operators of the
HQET.

The use of a hard (dimensionful) ultra-violet cut-off $\Lambda$ in the
effective theory (instead of dimensional regularisation) leads to
matrix elements and coefficient functions which are free of renormalon
ambiguities \cite{bigi}. However in this case the matrix elements in
the effective theory diverge as powers of the cut-off $\Lambda$. The
subtraction of these power divergences cannot be performed using
perturbation theory; such a subtraction reintroduces renormalon
ambiguities in the matrix elements and coefficient functions. This
point will be discussed in detail below.

In this paper, in contrast to previous approaches, we propose a
non-perturbative definition of the renormalised operators
$O^{HQET}_{n,\alpha}$, such that they are free of both renormalon
ambiguities and power divergences. This allows us to present
``physical" definitions  of parameters such as the binding energy
($\bar\Lambda$) or the kinetic energy ($-\lambda_1/2m_Q$) of the heavy
quark in a heavy hadron. By ``physical" we mean that they do not depend
on any regularisation or renormalisation scale, nor on the method used
to regulate the ultra-violet divergences (although they do depend on
the renormalisation prescription used to define them).

An immediate, and related, question concerns the choice of the mass
parameter $m_Q$ used in the expansion. The pole mass is ambiguous due
to the presence of infra-red renormalon singularities \cite{bb,bigi},
and so a different definition of $m_Q$ is required. We introduce the
definition of a ``subtracted pole mass", $m_Q^S$, from which the
renormalon ambiguities have been subtracted non-perturbatively.
$m_Q^S$ has the attractive property that it can be determined from the
mass of a hadron $H$ containing the heavy quark $Q$, by calculations
performed entirely within the HQET. For each such hadron $H$ we define
a parameter $\bar\Lambda\equiv m_H-m_Q^S$, and
$\bar\Lambda\rightarrow$ constant as $m_Q\rightarrow\infty$ (for
simplicity of notation we suppress the label $H$ on $\bar\Lambda$).
$\lb$  remains finite when the ultraviolet cut-off is taken to
infinity, and can  be computed numerically in lattice simulations of
the HQET. We discuss the question of the definition and evaluation of
$\lb$ to leading order in section \ref{sec:lambdabar}, and to
$O(1/m_Q)$ in section \ref{sec:lambda1}.

As a further illustration of our general procedure we study in section
\ref{sec:lambda1} the matrix elements which determine the kinetic
energy of the heavy quark in a hadron \footnote{Here we consider the
effective theory for a heavy quark at rest. The generalisation to an
arbitrary four-velocity is straightforward, although there are
considerable subtleties in formulating the effective theory at non-zero
velocity in Euclidean space \cite{ugo,mandula}.}
\beq \lambda _1 =
\frac{\langle H|\bar h \vec D^2 h |H\rangle}{2m_H}
\label{eq:lambda1}\eeq
where $h$ represents the heavy quark field in the HQET, and H is a hadron
(of mass $m_H$) containing a heavy quark (we suppress the label $H$ on
$\lambda_1$). The $\lambda_1$'s are important ingredients in the study
of the spectroscopy and inclusive decays of heavy hadrons. The
chromomagnetic operator $\bar h\sigma_{ij}G^{ij}h$, where $G_{ij}$ are
the spatial components of the gluon field strength tensor, appears at
the same order of the heavy quark expansion as the kinetic energy
operator.  However the spin structure of the chromomagnetic operator
ensures that its matrix elements are free of renormalon singularities.

In ref.\cite{mms}, together with L.Maiani, we pointed out that in the
lattice formulation of the HQET the evaluation of higher order terms in
$1/m_Q$ involves the appearance of ultra-violet singularities which
diverge as inverse powers of the lattice spacing $a$. This is due to
the fact that the higher order terms generally involve matrix elements
of higher dimensional operators which can mix with lower dimensional
ones (e.g. $\bar h \vec D^2 h$ can mix with $1/a \,\bar h D_4 h$ and
$1/a^2\,\bar h h$). In ref.\cite{mms} it was further stressed that
these divergences must be subtracted non-perturbatively, since factors
such as
\beq
\frac{1}{a}\exp\left(-\int^{g_0(a)}\frac{dg^{\,\prime}}{\beta(g^{\,\prime})}
\right)=\Lambda_{{\rm QCD}}
\label{eq:nonpert}\eeq
which do not appear in perturbation theory, give non-vanishing
contributions as $a\rightarrow 0$. Renormalons represent an explicit
example of non-perturbative effects of this kind. In this paper we
explain how the subtraction of power divergences and of
renormalon singularities, can be performed non-perturbatively,
illustrating our ideas with the evaluation of $\bar\Lambda$ and
$\lambda_1$.

Much of the discussion in this paper is presented within the framework
of lattice field theory, however all the theoretical questions
addressed below have to be faced with any ultra-violet regularisation
scheme \cite{mns}. Moreover, our proposed definitions of subtracted
higher dimensional operators in general, and of $\lb$ and $\lambda_1$
in particular, are in fact independent of the regularisation. The
primary aim of this paper is to provide an understanding of these
theoretical issues. In addition however, lattice simulations provide
the opportunity for the evaluation of the parameters of the HQET, such
as $\lb$ and $\lambda_1$, from first principles. In the following
sections we also explain how this can be done in principle, and in
ref.\cite{cgms} we present some results from an exploratory numerical
study (some preliminary numerical results, and a summary of the ideas
of this paper have been presented in ref.\cite{cgmsbielefeld}).
However, instead of using lattice simulations, it is also possible to
use other non-perturbative methods, such as QCD sum rules, to compute
the matrix elements of the operators defined by our prescription. These
matrix elements are free of renormalons and power divergences.

Throughout this paper we use the following notation for ultra-violet
cut-offs and renormalisation scales. We denote by $M$ the scale used to
define the renormalised operators in QCD ($O^{QCD}$), and by $\mu$ that
to define the operators in the HQET (the $O^{HQET}_{n,\alpha}$).  Thus
$\mu \ll m_Q \ll M$. It will also be convenient to consider bare
operators in the HQET, and we denote by $\Lambda$ the corresponding
ultra-violet cut-off. Where the discussion is particular to the lattice
formulation of the HQET we replace $\Lambda$ by $a^{-1}$.

The plan of the remainder of this paper is as follows. In the next
section we discuss the definition of $\bar\Lambda$, and explain how it
can be computed in lattice simulations. We explain our general
procedure for the definition of subtracted higher dimensional operators
in section \ref{sec:lambda1}, using the matrix elements $\lambda_1$ as
examples. Section \ref{sec:matching} contains a discussion of
``matching", i.e. the determination of the coefficient functions
$C_{n,\alpha}$ corresponding to our definition of the operators.
Finally in section \ref{sec:concs} we present our conclusions.

\section{The Binding Energy - $\bar\Lambda$} \label{sec:lambdabar}

In this section we discuss the definition of the heavy quark mass
$m_Q^S$, from which the ambiguities from the leading renormalon
singularities have been subtracted non-perturbatively. $m_Q^S$ can thus
be used as the expansion parameter in the HQET. We start however by
reviewing very briefly some of the relevant points concerning the
renormalon singularities in the propagator of the heavy quark.

\subsection{Renormalons in the Heavy Quark Propagator}
\label{subsec:bb}

Our discussion follows closely the presentation by Beneke and Braun
\cite{bb}, in which these authors study the heavy quark propagator in
the large $N_f$ limit, where $N_f$ is the number of light quark
flavours. We refer the reader to ref.\cite{bb} for more details, and
also to ref.\cite{mns} in which the discussion is extended to include
the terms of $O(1/m_Q)$ in the inverse propagator. To leading
non-trivial order in $1/N_f$, the renormalon singularities are obtained
by summing over an arbitrary number of light quark loops in the gluon
propagator. The Borel transformed gluon propagator (in the Landau
gauge) is written as
\beqn
D^{AB}_{\mu\nu}(k,u) & = & \sum_{n=0}^{\infty} \frac{1}{n!}\,
D^{AB}_{\mu\nu,n}(k)\,\left(\frac{-4\pi\,u}{\beta_0\,\alpha_s}\right)^n
\nonumber\\
 & = &
i\delta^{AB} \left(\frac{e^C}{\mu^2}\right)^{-u}
\frac{k_\mu k_\nu
- k^2 g_{\mu\nu}}{(-k^2)^{2+u}}
\label{eq:gluonborel}\eeqn
where $\beta_0$ is the coefficient of $\alpha_s/4\pi$ in the first term
of the $\beta$-function ($\beta_0 =-(11-2/3N_f)$\,), and $A$ and $B$
are colour labels. $\mu$ is the  renormalisation scale and $C$ is a
scheme dependent constant.  $D^{AB}_{\mu\nu,n}(k)$ is the contribution
to the gluon propagator from the diagram with $n$ quark loops. Of
course in leading order in  $N_f$ it is only the term $2/3N_f$ which
appears, however it is assumed that the replacement of $2/3 N_f$ by
$\beta_0$ is a consistent one for identifying the singularities. This
assumption is based on the intuition, that it is the infra-red
behaviour of the running coupling constant which is (at least
partially) responsible for the singularities.

Feynman diagrams evaluated with the Borel transformed propagator
(\ref{eq:gluonborel}) may have poles at positive values of $u$,
rendering the inverse Borel transform (which requires an integral over
positive values of $u$) ambiguous~\footnote{When the calculations are
extended beyond the leading order in $1/N_f$, the poles become replaced
by branch points of cut singularities.}. This is particularly true in
Operator Product Expansions, where such singularities in the Wilson
coefficient functions are cancelled by those in the matrix elements of
higher dimensional, or higher twist, operators
\cite{muelleraachen,mueller}. The expansion in inverse powers of the
mass of the heavy quark is an interesting example of this phenomenon.
We start by considering the quark propagator.

The inverse quark propagator in QCD can be written in the form
\beq
S^{-1}(p,m) = \slash \hspace{-7pt} p - m - \Sigma(p,m)
\label{eq:sinv}\eeq
where
\beq
\Sigma(p,m) = m \Sigma_1(p^2,m) + (\slash \hspace{-7pt} p -m)
\Sigma_2(p^2,m)
\label{eq:sigma}\eeq
and $m$ is the bare mass. We now write
\beq
p = m_Q v + k
\label{eq:ptok}\eeq
where $m_Q$ is some well defined choice of the heavy quark mass, to be
specified later. In all the explicit examples given in sections
\ref{subsec:lbdef} and \ref{sec:lambda1} we will take
$v=(1,\vec 0)$. It is also convenient to define the quark
propagator sandwiched between projection operators
\beq
\frac{1+\vslash}{2} S_P(k,m_Q) \equiv
\frac{1+\vslash}{2} S(p,m)\frac{1+\vslash}{2}
\label{eq:spdef}\eeq
Then, \cite{bb},
\beq
S_P^{-1}(k,m_Q) = m_Q - \mpole (m_Q,\mu) - C(m_Q/\mu)\,S^{-1}_
{{\rm eff}}(v\cdot k, \mu)
+ O\left(\frac{(v\cdot k)^2}{m_Q},\frac{k_\perp^2}{m_Q},
\frac{1}{N_f^2}\right)
\label{eq:spm1expansion}\eeq
where $S_{{\rm eff}}$ is the quark propagator in the HQET  (whose
action is given by $\bar h iv\cdot Dh$; $h$~represents the field of the
heavy quark), $\mu$ is the renormalisation scale, and
$k_\perp^2=k^2-(v\cdot k)^2$. $S^{-1}_{{\rm eff}}$ is linearly
divergent in perturbation theory. In dimensional regularisation the
Borel transforms of both $\mpole$ and $S^{-1}_{{\rm eff}}$ have
renormalon singularities at $u=1/2$, the infra-red renormalon in
$\mpole$ cancels the ultra-violet renormalon in $S^{-1} _{{\rm eff}}$
on the r.h.s. of eq.(\ref{eq:spm1expansion}) \cite{bb}. The connection
between power divergences and renormalons can be seen by noting that if
$S^{-1}_{{\rm eff}}$ is linearly divergent in one-loop perturbation
theory (which corresponds to using (\ref{eq:gluonborel}) with $u=0$ as
the gluon propagator), then its Borel transform is  logarithmically
divergent at $u=1/2$, corresponding to a pole singularity at this
point. In particular $S^{-1}_{{\rm eff}}(0, \mu )$ is not equal to
zero, which is a signal of the presence of ultra-violet renormalon
singularities at $u=1/2$. A more detailed account of the correspondence
between power divergences and renormalons is given in ref.\cite{mns}.

For the discussion below, it is convenient to re-express eq.(\ref
{eq:spm1expansion}) in terms of the bare propagator in the effective
theory with a hard cut-off $\Lambda$,
\beq
S_P^{-1}(k,m_Q) = m_Q - \mpole(m_Q,\Lambda) -
C(m_Q/\Lambda)\,S^{-1}_ {{\rm eff}}(v\cdot k, \Lambda)
+ O\left(\frac{(v\cdot k)^2}{m_Q},\frac{k_\perp^2}{m_Q},
\frac{1}{N_f^2}\right) \, .
\label{eq:spm1expansion2}\eeq
Now $\mpole(m_Q,\Lambda)$ and $S^{-1}_ {{\rm eff}}(v\cdot k, \Lambda)$
both diverge linearly with the cut-off, but their Borel transforms have
no poles at $u=1/2$. Thus we have replaced the problem of the
ambiguities associated with renormalon singularities with that of
determining the exponentially small terms when power divergences are
present (see eq.(\ref{eq:nonpert}) and the corresponding discussion).
By adding a residual mass counterterm  ($\delta m\,\bar hh$) to the
action of the HQET, it is possible to impose the non-perturbative
condition  $S^{-1}_{{\rm eff}}(0,\Lambda) =0$, thus removing the power
divergences. In the following subsection, we impose instead an
equivalent condition on the static (i.e. $\vec v=0$) heavy quark
propagator in Euclidean configuration space ($S^{ij}_{\rm eff}(x;y)$,
where $i$ and $j$ are colour labels) in the Landau gauge:
\beq
\lim_{t\rightarrow \infty}\,\frac{d}{dt}\ln\left(|S^{ii}_{\rm eff}
(x;0)|
\right)\, =0
\label{eq:condition}\eeq
where $t=x^0$ and $S^{ii}_{\rm eff}$ is the trace over the colour
components of the propagator. Since the operator we are subtracting
($\bar hh$) is conserved, either of these conditions is sufficient to
determine its coefficient $\delta m$ fully. Matching the effective
theory onto full QCD now implies that $\mpole$ on the r.h.s. of
eq.(\ref{eq:spm1expansion2}) is replaced by a ``subtracted pole mass",
$m_Q^S$, from which the power divergences (and indeed all the
dependence on $\Lambda$) have been subtracted. $m_Q^S$ is therefore a
natural choice for the expansion parameter $m_Q$ of the HQET. The
implementation of the above subtraction requires the non-perturbative
evaluation of the quark propagator in the HQET because of the arguments
given in the discussion of eq.(\ref{eq:nonpert}). Lattice simulations
provide the possibility for such an evaluation, and in the following
subsection we discuss an explicit definition of $m_Q^S$ based on
eq.(\ref{eq:condition}), and of the corresponding $\lb$ parameter,
$\lb=m_H-m_Q^S$, where $H$ is a hadron containing one heavy quark $Q$.
In spite of the fact that the entire discussion of the following
subsection is in the  framework of lattice field theory, the value of
$\lb$ is determined by  its definition based on the condition
(\ref{eq:condition}), and is independent of the method of
regularisation.

\subsection{Definition of $\lb$}
\label{subsec:lbdef}

In this subsection we present our definition of $\lb$ explicitly, and
also discuss how it might be evaluated using the lattice formulation of
the HQET.  The discussion is presented in Euclidean space, and we take
as the action of the HQET in the static (i.e. $\vec v=0$) case
\beq
{\cal L}_{\rm eff}= \bar h D_4h
\label{eq:action0}\eeq
where $h$ represents the field of the heavy quark. Consider the
correlation  function
\beq
C(t) = \sum_{\vec x}\langle 0|\,J_\Gamma(\vec x,t) \
\bar J^\dagger_\Gamma(\vec 0,0)\,|0\rangle
\label{eq:corrfun}\eeq
where $J^\dagger_\Gamma$ and $J_\Gamma$ are interpolating operators
which can create or annihilate a meson state in the HQET. For example,
we may take $J=\bar h\Gamma q$ where $q$ represents the  field of the
light quark and $\Gamma$ is one of the Dirac matrices \footnote{The
generalisation to heavy baryons is straightforward.}.  For sufficiently
large times, so that only the ground state contributes significantly to
the correlation function,
\beq
C(t) \rightarrow Z^2\exp(-{\cal E}t)
\label{eq:casymp}\eeq
where the constant $Z$ is independent of the time. The exponent ${\cal
E}$  is equivalent to the definition of $\bar\Lambda$ proposed in
ref.\cite{fln}
\beq
\bar\Lambda = \frac{-\partial_4\,\langle 0|\,\bar
h\Gamma q\,|M\rangle} {\langle 0|\,\bar h\Gamma q\,|M\rangle}
\label{eq:lambdabarfln}\eeq
However ${\cal E}$, and hence $\bar\Lambda$ defined through eq.(\ref
{eq:lambdabarfln}), is not a physical quantity, since it diverges
linearly as $a\rightarrow 0$ (as can be demonstrated in one-loop
perturbation theory). If the matrix elements in
eq.(\ref{eq:lambdabarfln}) are defined in the $\overline{MS}$ scheme,
then the corresponding definition of $\lb$ contains an ambiguity of
$O(\Lambda_{QCD})$. The linear divergence in ${\cal E}$ appears due to
the mixing of the operator $\bar hD_4 h$ in the effective action, with
the lower  dimensional operator $\bar hh$. It is possible to subtract
this  divergence non-perturbatively by adding a residual mass term
$\delta m\, \bar hh$ to the action, where $\delta m$ is determined by
imposing a suitable renormalisation condition. We now state explicitly
our proposal  for a physical definition of $\bar\Lambda$, and for its
non-perturbative evaluation:

\begin{enumerate}

\item[i)] Evaluate the heavy quark propagator in the theory defined by
the action (\ref{eq:action0}), in some fixed gauge (the Landau gauge
say). In practice this will be done using lattice simulations. The
heavy quark propagator in a smooth gauge, such as the Landau gauge, is
of the form (for $t>0$)
\beq
S_h^{ij}(\vec x,t;\vec 0,0) = \langle\, 0\,|\, h^i(x)\bar h^j (0)\,|\,0\,
\rangle=\delta^3(\vec x)\,\delta^{ij} A(t) \exp (-\lambda t)
\label{eq:shform}\eeq
where the ultra-violet divergences associated with the quark mass are
contained in the exponent $\lambda$. $i$ and $j$ are colour labels, and
unless specifically required they will be suppressed below. $A(t)$
satisfies the condition
\beq
\lim_{t\rightarrow\infty}\,\frac{d}{dt} \ln\{|A(t)|\} = 0
\label{eq:asmooth}\eeq

In the explicit examples below we will use the following lattice covariant
derivative
\beq
D_4f(\vec x,t) = \frac{1}{a}\left( f(\vec x,t) -
U^\dagger_4(\vec x, t-a)f(\vec x, t-a)\right)
\label{eq:d4def}\eeq
where $\{U_\mu(\vec x,t)\}$ are the link variables. Other lattice
definitions of $D_4$ are also acceptable.
\item[ii)] Add a residual mass term to the effective action
\footnote{The normalisation factor $1/(1 + \delta m\,a)$ is introduced
for convenience, as will become apparent below.}
\beq
{\cal L}^\prime_{{\rm eff}} = \frac{1}{1 + \delta m\,a}\left(
\bar hD_4h + \delta m\, \bar hh\right)
\label{eq:lprime}\eeq
where $\delta m$ will be specified in iii) below. With the new action,
the heavy quark propagator (now denoted by $S_h^\prime$) is given by
\beq
S^\prime_h(\vec x, t;\vec 0,0) =
S_h(\vec x, t;\vec 0,0) \,\exp\!\left(-\ln(1 + \delta m\,a)t/a\right)
\label{eq:shprime}\eeq
This result demonstrates that the divergences associated with mass
renormalisation do indeed exponentiate.

\item[iii)] We fix the residual mass counterterm $\delta m$ by the
condition
\beq
-\nu = \frac{\ln (1 + \delta m\,a)}{a}\equiv
\lim_{t\rightarrow\infty}\,\frac{1}{a} \ln\!\left(\frac{|S^{ii}_h(\vec x, t +
a;\vec 0,0)|}{|S_h^{ii}(\vec x, t;\vec 0,0)|}\right)
\label{eq:deltamdef}\eeq
$\nu$ can be computed in numerical simulations by studying the
logarithm on the right hand side of eq.(\ref{eq:deltamdef}) as a
function of $t$.

\item[iv)] The ``physical" definition of $\lb$ is
\beq
\bar\Lambda \equiv {\cal E} - \nu = {\cal E} + \frac{1}{a}\,\ln(1\, +\,
\delta m\, a)
\label{eq:lambdabardef}\eeq
\underline{$\lb$ is finite in the limit $a\rightarrow 0$ and is free
of renormalon ambiguities}.
The corresponding ``subtracted" pole mass $m_Q^S$ is defined by
\beq
m_Q^S \equiv m_H - \lb\ .
\label{eq:mqsdef}\eeq
$m_Q^S$ is independent of $a$, and contains no renormalon ambiguities.
Thus from the computed value of $m_Q^S$ one can determine \mmsbar
$(\mu)$, or any other short-distance definition of the quark mass (up
to uncertainties which are now of $O(1/m_Q)$), using perturbation
theory. This will be discussed in section \ref{sec:matching}.

The counterterm $\nu$ defined above is gauge-invariant, in spite of the
fact that it is calculated from the heavy quark propagator in a fixed
gauge. The argument goes as follows. The linear divergence is
eliminated from any correlation function, i.e. for any external state,
by subtracting  from the action (\ref{eq:action0}) a term proportional
to the gauge-invariant operator $\bar h h$. Since in this way one
eliminates all divergences both for quark and hadron  external states,
the coefficient of the mixing has to be gauge-invariant. This must be
true also for the finite non-perturbative term which accompanies the
linear divergence. The gauge-invariance of the linearly divergent term
has been checked  explicitly in one-loop perturbation theory. The same
argument can be applied to the evaluation of the subtraction constants
which appear in the definition of a finite kinetic energy operator, to
be discussed in the following section.
\end{enumerate}

It may appear more natural to define $\lb$ using
eqs.(\ref{eq:deltamdef}) and (\ref{eq:lambdabardef}), but with $\nu$
determined at a small value of $t$, ($t^*$ say, with $1/t^*\gg
\Lambda_{QCD}$). In particular it may seem that the value of $\lb$
obtained from a measurement at ``short distances" can be used more
reliably to determine some standard short distance mass in QCD
(\mmsbar$(\mu)$ say, with $\mu \gg \Lambda_{QCD}$) using perturbation
theory. This is not the case however, since in addition to the
non-perturbative contribution of $O(\Lambda_{QCD})$ to $\lb$, there is
a perturbative contribution which is proportional to $1/t^*$,
\beqn
-\nu_{\rm pert}(t^*) &\equiv & \frac{1}{a}\,\ln\left(\frac{|S^{ii}_{\rm
pert}(\vec x,t^*+a;\vec 0,0)|} {|S^{ii}_{\rm pert}(\vec x,t^*;\vec
0,0)|}\right)\nonumber\\
& = & -\frac{1}{a}\, \frac{\alpha_sC_F}{4\pi}\,\gamma_\psi\, \ln\left(1 +
\frac{a}{t^*}\right)\ +\  O(\alpha_s^2) \\
& \rightarrow & -
\frac{\alpha_sC_F}{4\pi} \,\frac{\gamma_\psi}{t^*} \ +\  O(\alpha_s^2)
\ \ \ \ \ \ \ \left({\rm as}\ \ a\rightarrow 0\right) \eeqn
where $S_{\rm pert}$ is the heavy quark propagator in perturbation
theory, which for $t>0$ takes the form
\beq
S^{ij}_{\rm pert} (\vec x,t;\vec 0,0) =
\delta^{(3)}(\vec x)\,\delta^{ij} \left[ 1 -  \frac{\alpha_sC_F}{4\pi}
\left(\gamma_\psi \ln (t/a) + c_t\right) + ... \right]
\label{eq:stpert}\eeq
and the anomalous dimension of the heavy quark field ($\gamma_\psi$)
and $c_t$ are constants (in the Landau gauge $\gamma_\psi =-6$). $C_F$
is the eigenvalue of the quadratic Casimir operator in the fundamental
representation ($C_F$=4/3).  In order to determine $\lb$, from the
propagator computed at a finite value of $t^*$, the perturbative
contribution must be subtracted. The evaluation of the term
proportional to $1/t^*$ in perturbation theory (which in practice can
only be performed up to some low order), becomes less accurate as $t^*$
is decreased.   The reason is that, although the calculation of the
coefficient of the term proportional to $1/t^*$ becomes more accurate
as $t^*$ decreases, the presence of the factor $1/t^*$ implies that the
subtraction becomes larger numerically and that the error due to
(unknown) higher order pertubative corrections also increases, reducing
the accuracy of the result for $\lb$. For this reason we propose to
define $\delta m$ from eq.(\ref{eq:deltamdef}), i.e. from measurements
of the propagator at large values of $t$. In some simulations it may
not be possible to compute the propagators at sufficiently large values
of $t$ for a plateau to be reached (i.e. for the ratio of the
propagators on the right hand side of eq.(\ref{eq:deltamdef}) to be
independent of $t$). In those cases it may be necessary to determine
$m_Q^S$ from measurements taken at intermediate values of $t$,  and to
perform the subtraction of the terms proportional to $1/t$,  either by
using perturbation theory or by fitting $-\nu$ to a function of  $t$
and extracting the asymptotic value (i.e. the value as
$t\rightarrow\infty$). In refs.\cite{cgms,cgmsbielefeld} it has been
shown that the latter method can be used to give a precise
determination of $\lb$, and hence of $m_Q^S$.

\section{The Kinetic Energy - $\lambda_1$}\label{sec:lambda1}

In this section we present our proposal for the elimination of power
divergences and renormalons from the matrix elements of higher
dimensional operators. We illustrate our method by considering
explicitly the kinetic energy operator $\bar h \vec D^2 h$. The matrix
elements of this operator contain power divergences because it can mix
with the lower dimensional operators $\bar hD_4 h$ and $\bar
hh~$\footnote{ The relation between power divergences in matrix
elements of the particular operator $\bar h \vec D^2 h$ and renormalon
ambiguities in dimensional regularisation is subtle and not fully
understood. A detailed discussion of this subject can be found in ref.
\cite{mns}.}. A subtracted kinetic energy operator, one which is free
of power divergences, is of the form
\beq
\bar h \vec D^2_S h = \bar h \vec D^2 h - c_1 (\bar h
D_4 h + \delta m \bar hh) - c_2 \bar hh
\label{eq:d2sdef}\eeq
where the constants $c_1$ and $c_2$ are fixed by imposing appropriate
renormalisation conditions. We propose to define $c_1$ and $c_2$  by
imposing that the matrix element of $\bar h \vec D^2_S h$ between quark
states with $\vec k=0$ (where $k$ is the momentum of the quark), and in
the Landau gauge, vanishes
\beq
\langle h(\vec k=0)|\bar h \vec D^2_S h | h(\vec k=0)\rangle = 0\, .
\label{eq:subtraction}\eeq
Although not unique, this is perhaps the most intuitive definition
of the kinetic energy of the heavy quark in a hadron.
Specifically we determine the constants $c_1$ and $c_2$ by using~\footnote
{From now on we will work in lattice units, setting $a=1$.}
\beq
c_1 + c_2
t_x = \frac{\sum_{\vec x,\vec y}\sum_{t_y =0,t_x}
\langle\, 0\,|\, h(\vec x,t_x)\ \bar h(\vec y,t_y)\vec D^2_y h(\vec y,t_y)\
\bar h(\vec 0,0)\,|\,0\,\rangle}
{\sum_{\vec x}\,\langle\,0\,|\, h(\vec x,t_x)\bar h(\vec 0,0)\,
|\,0\,\rangle}\, .
\label{eq:c1c2}\eeq
In deriving eq.(\ref{eq:c1c2}) it is implied that we are using the
action ${\cal L}^\prime$ given in eq.(\ref{eq:lprime}) which contains
the residual mass counterterm. However, from the discussion in the
previous section we can readily see that this is equivalent to using
the action ${\cal L}$ given in eq.(\ref{eq:action0}) which has no
residual mass term. The only difference in using the actions ${\cal
L}^\prime$ and ${\cal L}$ is a factor  $\exp (\nu t_x)$ in both the
numerator and denominator of the right hand side of eq.(\ref{eq:c1c2}).

For some important applications it is only the constant $c_2$ which is
required. This is because
\beq
\frac{1}{1+\delta m}(\bar h D_4 h + \delta m \bar
hh)S_h^\prime(x;y)=\delta^{4}(x-y)\, ,
\label{eq:dirac}\eeq
so that if the operators in a correlation function are separated (i.e.
up to contact terms), then the term proportional to $c_1$ vanishes.
$c_2$ can also be determined  directly by eliminating the sum over
$t_y$ in eq.(\ref{eq:c1c2}):
\beq
c_2 = \frac{\sum_{\vec x,\vec y}\,
\langle\, 0\,|\, h(\vec x,t_x)\ \bar h(\vec y,t_y)\vec D^2_y h(\vec y,t_y)\
\bar h(\vec 0,0)\,|\,0\,\rangle}
{\sum_{\vec x}\,\langle\,0\,|\, h(\vec x,t_x)\bar h(\vec 0,0)\,
|\,0\,\rangle}\, ,
\label{eq:c2}\eeq
for $t_y\neq 0, t_x$.

Having defined the subtracted operator $\bar h\vec D_s^2h$, $\lambda_1$ can
be determined from a computation of two- and three-point correlation
functions in the standard way. Consider the meson three-point
correlation function (the extension of this discussion to baryons
is entirely straightforward)
\beq
C_{\vec D^2_S}(t_x,t_y) = \sum_{\vec x}
\langle 0|J_\Gamma(\vec x, t_x)\ \bar h(\vec y,t_y)\vec D^2_S h(\vec
y,t_y)
J^\dagger_\Gamma(\vec 0,0)|0\rangle
\label{eq:cd2s}\eeq
For sufficiently large values of $t_y$ and $t_x-t_y$
\beq
C_{\vec D^2_S}(t_x,t_y)\ \rightarrow Z^2\lambda_1\exp
\left(-({\cal E}-\nu)t_x\right)
\label{eq:cd2sasymp}\eeq
where $\lambda_1$ is defined by eq.(\ref{eq:lambda1}), using  the
subtracted kinetic energy operator,
\beq
\lambda_1=\frac{\langle H|\bar h\vec D^2_S h|H\rangle}{2m_H} \, ,
\label{eq:deltadef}\eeq
and $H$ is the lightest meson state which can be created by the
operator $J_\Gamma^\dagger$. A convenient way to extract $\lambda_1$ is to
consider the ratio
\beq
R(t_x,t_y)=\frac{C_{\vec D^2_S}(t_x,t_y)}{C(t_x)}\rightarrow\lambda_1
\label{eq:deltaasymp}\eeq
As usual $\lambda_1$ must be evaluated in an interval in which
$R(t_x,t_y)$ is independent of the times $t_y$ and $t_x$, so that the
contribution from the excited states can be neglected. By the same
argument as was given after eq.(\ref{eq:c1c2}), the ratio in
eq.(\ref{eq:deltaasymp}) can be evaluated using the action ${\cal L}$
of eq.(\ref{eq:action0}) with no residual mass term. In the present
case the difference between the matrix elements of the subtracted and
unsubtracted operators is given by
\beq
\lambda_1\equiv \frac{\langle H|\bar h\vec D^2_S h|H\rangle}{2m_H}
= \frac{\langle H|\bar h\vec D^2 h|H\rangle}{2m_H}\, - \, c_2\, .
\label{eq:deltau}\eeq

We conclude this section by presenting the definition of the subtracted
quark mass up to, and including the terms of $O(1/m_Q)$,
\beq
\tilde M_B = m_Q^S + {\cal E} -\nu - \frac{\lambda_1}{2m_Q^S}
\label{eq:mqnext}\eeq
where $\tilde M_B$ is the spin-averaged mass, $\tilde M_B = \frac{1}{4}
(M_B+3M_{B^*})$, which has no contribution from the chromomagnetic
operator. Eq.(\ref{eq:mqnext}) must be modified to include the effects of
perturbative corrections. We denote the renormalised kinetic energy
operator in some continuum renormalisation scheme by  $\bar hD^2_{{\rm
cont}}h$. In one loop perturbation theory we have
\beq
\bar hD^2_{{\rm cont}}h = \left( 1+ \frac{\alpha_s}{4\pi}X_{\vec
D^2_S}\right)
\bar hD^2_{S}h
\label{eq:d2cont}\eeq
from which we derive
\beq
\tilde M_B = m_Q^S + {\cal E} -\nu - \left( 1+
\frac{\alpha_s}{4\pi}X_{\vec D^2_S}\right)\frac{\lambda_1}{2m_Q^S}
\label{eq:mqnextrad}\eeq
The term proportional to $X_{\vec D^2_S}$ in eq.(\ref{eq:mqnextrad}) is
absent in continuum formulations of the HQET, and is a manifestation
of the lack of reparametrisation invariance in the lattice version. It
has been calculated in ref.\cite{mms}.

\par Although we have restricted our explicit discussion to the kinetic
energy operator, clearly the same techniques can be applied to a wide
class of operators. This includes, for example, the operators whose
matrix elements determine the $1/m_Q$ corrections to exclusive leptonic
and semileptonic decays of $B$-mesons \cite{neubert}. In each case one
can construct linear combinations of higher and lower dimensional
operators which are free of power divergences and renormalon
ambiguities, by  imposing appropriate normalisation conditions for
matrix elements between quark states in a fixed gauge, and at given
momenta. This  approach is a particular application of the general
method for the non-perturbative normalisation of lattice operators
proposed in ref.\cite{mpstv}.

\section{Matching}\label{sec:matching}

In the preceeding sections we have proposed a method for defining
higher-dimensional operators $O_{n,\alpha}^{{\rm HQET}}$, whose matrix
elements are free of (ultra-violet) renormalon ambiguities and power
divergences, and which can be computed in lattice simulations. In order
to derive physical predictions from these matrix elements it is
necessary to calculate the corresponding coefficient functions, i.e.
the $C_{n,\alpha}$ of eq.(\ref{eq:ope}). The subtraction of the
ultra-violet renormalons from the matrix elements, implies the
elimination of the corresponding infra-red renormalons from the
coefficient functions, which can therefore be computed in perturbation
theory \cite{mueller,grunberg,mnflow}. To illustrate the ``matching" procedure
consider a simple situation for which we can write eq.(\ref{eq:ope})
as:
\beq
O^{QCD} (M) = C_1(\frac{m_Q}{M}, \frac{\Lambda}{M}) O_1^{HQET}(\Lambda) +
\frac{1}{m_Q}C_2(\frac{m_Q}{M}, \frac{\Lambda}{M}) O_2^{HQET}(\Lambda) +
O(1/m_Q^2)
\label{eq:opeexample}\eeq
i.e. where there is a single operator in each of the first two terms of
the heavy quark expansion (the discussion can readily be extended to
the general case). In eq.(\ref{eq:opeexample}) we have exhibited
explicitly the dependence on the scale $M$ used to normalise the QCD
operators  $O^{QCD}(M)$ (i.e. the operators in the full theory).
$\Lambda$ is the (hard) ultra-violet cut-off in the  effective theory
(for example in the lattice formulation of the HQET  $\Lambda =
a^{-1}$), and the operators on the right-hand side of
eq.(\ref{eq:opeexample}) are bare operators.  $m_Q$ can be considered
as the subtracted pole mass $m_Q^S$ defined in section
\ref{sec:lambdabar}.

Note that although we have chosen to write eq.(\ref{eq:opeexample}) in
terms of the bare operators in the effective theory, this equation is
equivalent to eq.(\ref{eq:ope}), which is written using renormalised
operators. From the bare operators (computed in lattice simulations for
example), one can determine the corresponding ones renormalised in a
given (continuum) scheme by using perturbation theory, or by some
non-perturbative method (such as that given in ref.\cite{mpstv}). The
determination of the  matrix elements of renormalised operators of the
effective theory may be a convenient intermediate step, but it is not
necessary. We choose instead to compute the matrix elements of the
$QCD$ operators $O^{QCD}$ directly from the bare operators of the
effective theory.

With a hard cut-off the Borel transforms of the coefficient function
$C_1$ and the matrix elements of $O_2^{HQET}$ do not have renormalon
singularities at $u=1/2$. For example, in the large $N_f$ limit, the
Borel transform of $C_1$ in the vicinity of $u=1/2$ will have the
structure
\beq
\tilde C_1(\frac{m_Q}{M}, \frac{\Lambda}{M}, u)\propto \frac{1}{(1-2u)}
\left[\left(\frac{m_Q}{M}\right)^{-2u}-
\frac{\Lambda}{m_Q}\left(\frac{\Lambda}{M}\right)^{-2u}\right]\,
\label{eq:example}\eeq
where the tilde denotes the Borel transform. Although the residue of
the pole vanishes at $u=1/2$, both $C_1$ and the matrix elements of
$O_2^{HQET}$ contain terms which, in perturbation theory, diverge
linearly with the ultra-violet cut-off $\Lambda$. The use of a hard
factorisation scale to organise the heavy quark expansion has been
suggested by Bigi et al. \cite{bigi} (see also ref.\cite{dis} for a
very recent study of the definition of the higher twist operators
relevant for studies of deep inelastic structure functions). Our
proposal is to subtract these power divergences non-perturbatively, and
to use the subtracted operators, which are free of power divergences
and renormalon ambiguities, as the basis for the expansion in
eq.(\ref{eq:ope}).

Following the discussion in the preceding sections, we define a
subtracted operator $O_{2;S}^{HQET}$ which does not mix with
$O_1^{HQET}$,
\beq
O_{2;S}^{HQET}(\Lambda) = O_2^{HQET}(\Lambda) -
c(\Lambda)O_1^{HQET}(\Lambda)
\label{eq:o2sdef}\eeq
where the dependence of the matrix elements of
$O_{2;S}^{HQET}(\Lambda)$  on $\Lambda$ is at most logarithmic.
$c(\Lambda)$ is computed non-perturbatively, whereas the functions
$C_1$ and $C_2$ are calculated in perturbation theory.
Eq.(\ref{eq:opeexample}) can be rewritten as:
\beqn
O^{QCD}(M) & = & \left[C_1(\frac{m_Q}{M}, \frac{\Lambda}{M})
+ \frac{c(\Lambda)}{m_Q}C_2(\frac{m_Q}{M}, \frac{\Lambda}{M})\right]
O_1^{HQET}(\Lambda)  \nonumber\\
& &
\mbox{\hspace{0.4in}}+ \frac{1}{m_Q}C_2(\frac{m_Q}{M},
\frac{\Lambda}{M})\, O_{2;S}^{HQET}(\Lambda) +
O(1/m_Q^2)
\label{eq:opesexample}\eeqn
By using a combination of perturbative and non-perturbative techniques,
we have ensured that the coefficient of $O_1^{HQET}$ in
eq.(\ref{eq:opesexample}) does not contain terms which diverge linearly
with $\Lambda$, nor ambiguities due to renormalon singularities at
$u=1/2$. The remaining ambiguities are of $O(\Lambda_{QCD}^2/m_Q^2)$ or
less, and are associated with the renormalons at $u=1,
\frac{3}{2}\cdots$, or the corresponding power divergences in matrix
elements of operators of higher dimension than $O_2^{HQET}$. These can
be eliminated by generalising our procedure to higher orders of the
heavy quark expansion, by defining subtracted operators
$O_{3;S}^{HQET}$, $O_{4;S}^{HQET}\,\cdots$ which cannot mix with lower
dimensional ones, and using these subtracted operators as the basis of
the expansion in eq.(\ref{eq:ope}). Indeed by using such a basis one
eliminates the ambiguities from all coefficient functions, up to the
order in $1/m_Q$ for which the subtraction coefficients have been
computed. Of course in general the coefficient function of $O_1^{HQET}$
depends logarithmically on $\Lambda$, the dependence being given by the
anomalous dimension of $O_1^{HQET}$.

Throughout this paper we have been stressing the necessity of
performing the subtractions of power divergences non-perturbatively. We
now present a specific example demonstrating this explicitly.  Consider
the evaluation of the subtraction constant $c(\Lambda)$ in the case for
which $O_1^{HQET}=\bar hh$ and $O_2^{HQET}=\bar h\,v\cdot D\,h$ as
happens in the expansion of the  QCD operator $\bar QQ$, where $Q$ is
the field of the heavy quark in the full theory,  and in the discussion
of $\lb$ in section \ref{subsec:lbdef}. The Borel transform of the
forward matrix element of $O_2^{HQET}$ between heavy quark states of
momentum $k$  ($T=\langle h(k)|\,\bar h v\cdot D  h\, | h(k)\rangle$),
in the  vicinity of $u=1/2$, has the structure
\beq
\tilde T(u) \propto \frac{1}{(1-2u)} \left[v\cdot k\left(\frac{v\cdot
k}{\Lambda}\right)^{-2u}\, -\,  \Lambda\,\right]\, .
\label{eq:example2}\eeq
The residue of the pole vanishes at $u=1/2$, in a similar way to that
in the coefficient function $C_1$ in eq.(\ref{eq:example}).
The two terms in the square brackets in eq.(\ref{eq:example2}),
contribute equal and opposite ambiguous terms of $O(\Lambda_{QCD})$ to
the matrix element of $O_2^{HQET}$ at $v\cdot k=0$. In perturbation
theory however, the first term does not appear at $v\cdot k=0$, and
hence the Borel transform of $c(\Lambda)$ contains a renormalon at
$u=1/2$.  It is for this reason that we insist on the non-perturbative
determination of the subtraction coefficients.

\par One can also imagine performing the matching completely
non-perturbatively, by simulating both the HQET and QCD (with a bottom
quark) on the lattice. However the latter requires a very small lattice
spacing, $a^{-1}\gg m_b$, which will not be possible for some time to
come. Moreover, once one is able to simulate the $b$-quark directly
(and reliably) on the lattice, the necessity of using the HQET is
removed. It may still however, be a useful guide to scaling properties
and symmetry relations.

\section{Conclusions}\label{sec:concs}

In this paper we have presented a method for defining higher
dimensional operators of the HQET, in such a way that their matrix
elements are free of ambiguities due to (ultra-violet) renormalon
singularities, and of power divergences. We have illustrated our
approach by proposing a physical definition of $\lb$ ($=m_H-m_Q^S$) and
of the matrix elements of the subtracted kinetic energy operator
($\langle H|\vec D^2_S|H\rangle$). The definition of the higher
dimensional operators involves the subtraction of lower dimensional
ones with the same quantum numbers. The subtraction coefficients are
determined by imposing normalisation conditions on Green functions
between quark and gluon states (or in the case of $\lb$, on the heavy
quark propagator). Lattice simulations of the HQET allow for a
numerical evaluation of the subtraction coefficients, as well as of the
matrix elements of the subtracted operators. The renormalisation
procedure proposed above can, however, be applied with any other
non-perturbative method for computing matrix elements in effective
theories. Our approach can also be extended to other cases, for example
to the higher twist contributions to the structure functions of deep
inelastic scattering.

\par Having defined operators $O_{n,\alpha}^{HQET}$ whose matrix
elements are free of renormalon ambiguities, we still have to match the
HQET operators onto those of QCD, i.e. to determine the  coefficient
functions $C_{n,\alpha}$ of eq.(\ref{eq:ope}). This is done using
perturbation theory, by calculating the matrix elements of $O^{QCD}$
and $O^{HQET}_{n,\alpha}$ between suitable external states, and
combining the results with the subtraction coefficients which have been
computed non-perturbatively (see section \ref{sec:matching} above).

\par The procedures defined in this paper allow one to study
quantitatively many important physical processes and quantities in
heavy quark physics using a systematic expansion in the mass of the
heavy quark. These include the leptonic and semi-leptonic decays of
heavy mesons and baryons, as well as relations between the masses and
lifetimes of heavy hadrons. The ``subtracted pole mass" is a suitable
parameter for the expansion.

\section*{Acknowledgements}

We thank M.Beneke, M,Crisafulli, V.Gimenez, M.Neubert and J.Nieves for
interesting discussions.  G.M. acknowledges partial support from
M.U.R.S.T., and CTS acknowledges the Particle Physics and Astronomy
Research Council for its support through the award of a Senior
Fellowship. We also acknowledge partial support by the EC contract
CHRX-CT92-0051.


\begin{thebibliography}{999}
\bibitem{nw} S.Nussinov and W.Wetzel, \pr {D36} (1987) 130
\bibitem{sv} M.B.Voloshin and M.A.Shifman, Sov.J.Nucl.Phys.
\underline{45} (1987) 292; \\
Sov.J.Nucl.Phys. \underline{47} (1988) 511
\bibitem{iw} N.Isgur and M.B.Wise, \pl{B232} (1989) 113; \\
\pl{B237} (1990) 527
\bibitem{georgi} H.Georgi, \pl{B240} (1990) 447
\bibitem{neubert} M.Neubert \prep{245} (1994) 259
\bibitem{muelleraachen} A.H.Mueller in ``QCD - 20 Years Later", eds.
P.M.Zerwas and H.A.Kastrup (World Scientific, Singapore 1993) p.162
\bibitem{bb} M.Beneke and V.M.Braun, \np{B426} (1994) 301
\bibitem{bigi} I.I.Bigi, M.A.Shifman, N.G.Uraltsev and A.I.Vaishtein,
\pr {D50} (1994) 2234
\bibitem{ns} M.Neubert and C.T.Sachrajda, CERN preprint TH.7312/94 (1994)
\\ (hep-ph/9407394)
\bibitem{manohar} M.Luke, A.V.Manohar and M.J.Savage, University
of Toronto preprint UTPT-94-21 (1994) (hep-ph/9407407)
\bibitem{ugo} U.Aglietti, \np{B421} (1994) 191;
U.Aglietti, M.Crisafulli and M.Masetti, \pl{B294} (1992) 281;
U. Aglietti and S.Capitani, Rome Preprint 993-1994
(hep-ph/9401335)
\bibitem{mandula} J.E.Mandula and M.C.Ogilvie, \pr{D45} (1992) 2183
\bibitem{mms} L.Maiani, G.Martinelli and C.T.Sachrajda, \np{B368}
(1992) 281
\bibitem{mns} G.Martinelli, M.Neubert and C.T.Sachrajda,
CERN Preprint 7540/94 (1995) in preparation
\bibitem{cgms} M.Crisafulli, V.Gimenez, G.Martinelli and C.T.Sachrajda,
University of Rome Preprint 94/1071 (1994)
\bibitem{cgmsbielefeld} M.Crisafulli, V.Gimenez, G.Martinelli and
C.T.Sachrajda, University of Rome Preprint (1994) (hep-lat/9412049)
\bibitem{mueller} A.H.Mueller, \pl{B308} (1993) 355
\bibitem{fln} A.F.Falk, M.Neubert and M.Luke, \np{B388} (1992) 363
\bibitem{mpstv} G.Martinelli, C.Pittori, C.T.Sachrajda, M.Testa and
A.Vladikas, Rome preprint 94/1022 (hep-lat/9411010)
\bibitem{grunberg} G.Grunberg, \pl{B325} (1994) 441
\bibitem{mnflow} M.Neubert, CERN Preprint 7487/94 (1994) (hep-ph/9412265)
\bibitem{dis} X.Ji, MIT preprint MIT-CTP-2381 (1994) (hep-ph/9411312)
\end{thebibliography}
\end{document}